\documentclass{PoS}

\usepackage{braket}
\usepackage{amsmath}
\usepackage{wrapfig }

\newcommand{\operator}[1]{\ensuremath{\hat{#1}}}
\newcommand{\openone}{\ensuremath{\, 1\hspace{-1ex}1}}

\title{Applying the variational principle to $(1+1)$ dimensional relativistic quantum field theories}

\ShortTitle{Applying the variational Principle to relativistic QFTs}

\author{\speaker{Jutho Haegeman}\\
        UGent, Department of Physics and Astronomy, Krijgslaan 281 S9, B-9000 Gent, Belgium\\
        E-mail: \email{jutho.haegeman@ugent.be}}

\author{J.~Ignacio Cirac\\
        Max-Planck-Institut f\"ur Quantenoptik, Hans-Kopfermann-Str. 1, Garching, D-85748, Germany}
\author{Tobias J. Osborne\\
        Wissenschaftskolleg zu Berlin, Berlin D-14193, Germany\\
       Leibniz Universit\"at Hannover, Institut f\"ur Theoretische Physik, Appelstr. 2, D-30167 Hannover, Germany}
\author{Henri Verschelde\\
        UGent, Department of Physics and Astronomy, Krijgslaan 281 S9, B-9000 Gent, Belgium}
\author{Frank Verstraete\\
        University of Vienna, Faculty of Physics, Boltzmanngasse 5, A-1090 Wien, Austria}

\abstract{We extend the recently introduced continuous matrix product state (cMPS) variational class to the setting of $(1+1)$-dimensional relativistic quantum field theories. This allows one to overcome the difficulties highlighted by Feynman concerning the variational procedure applied to relativistic theories, and provides a new way to regularize quantum field theories. A fermionic version of the continuous matrix product state is introduced which is manifestly free from fermion doubling and sign problems. We illustrate the power of the formalism with the simulation of free massive Dirac fermions, the Gross-Neveu model, and the Casimir effect. We find that cMPS can capture chiral symmetry breaking with absolute scaling of the chiral parameter, and that boundary effects can be accommodated with modest computational effort.}

\FullConference{The many faces of QCD\\
		November 1-5, 2010\\
		Gent, Belgium}

\begin{document}

\section{Introduction: Feynman's arguments against the variational principle}
The variational principle asserts that for any state $\ket{\Psi}$ in the Hilbert space $\mathbb{H}$ of a system with Hamiltonian $\operator{H}$ one finds an energy expectation value that exceeds the ground state energy, \textit{i.e.} 
\begin{equation*}
E_{0} \leq \frac{\braket{\Psi|\operator{H}|\Psi}}{\braket{\Psi|\Psi}},
\end{equation*}
with $E_{0}$ the ground-state energy (lowest eigenvalue) of $\operator{H}$. If we have a class of variational ansatz states $\ket{\Psi(z)}$ which are parameterized by a set of parameters $z$, we can try to find a good approximation of the ground state of $\operator{H}$ within this variational class by finding the parameters $z^{\ast}$ that minimize the energy expectation value. The variational method is the basis for a tremendous number of highly successful calculational tools in many-body physics. Examples include density functional theory \cite{dft}, Wilson's numerical renormalisation group \cite{Wilson} and the density matrix renormalisation group (DMRG) \cite{White}. When applicable, the variational method offers some powerful advantages over alternative approaches. It is free of any sign problem that hinders the application of Monte-Carlo sampling to many interesting problems, and it is perfectly able to reproduce non-perturbative effects and thus easily outperforms perturbation theory. 

In quantum field theory (QFT), however, the variational principle has not met with the same success as in other areas of many-body physics. The core reasons for this were identified by Feynman in one of his last lectures \cite{Feynman}. Feynman pointed out three conceptual issues standing in the way of a successful application of the variational principle in (relativistic) QFTs. Feynman phrased his first argument as the ``sensitivity to high frequencies''. This sensitivity is intrinsic to the variational method and its attempt to find the ``lowest'' ground-state energy. While this problem occurs in any system containing a large range of interacting energy scales, it is truly catastrophic in relativistic field theories. To lowest order, the ground state of a QFT contains the zero-point fluctuations from all energy scales and the corresponding ground-state energy is thus dominated by the contribution of the high frequencies. In relativistic theories, these UV frequency modes are infinitely abundant and create a divergence in the ground state energy (density), which already signals a difficulty that the variational principle will face. Any variational approach exploits all variational parameters $z$ to obtain the best possible description of the UV degrees of freedom and doesn't care about the relatively tiny energy penalty resulting from having an ill-described low frequency behavior. Since quantities of physical interest are related to the low-frequency modes, they will typically be very badly approximated when using the variationally optimized state $\ket{\Psi(z^{\ast})}$. Whenever the variational parameters affect both the low and high frequencies --- unavoidable in interacting theories but also true when a real space approach is applied to free theories --- this argument can lead to the paradoxical situation where the addition of variational parameters provides a \emph{worse} approximation to physical quantities.

Feynman's second and third arguments concern the lack of suitable variational ansatz states allowing high-accuracy computations of observable quantities, rather than an inherent problem of the variational method. As a second argument, Feynman remarks that a suitable ansatz for an extended quantum field theory should be extensive, \textit{i.e.} the energy expection value of the Hamiltonian with respect to a trial state $\ket{\Psi}$ should be proportional to the volume $V$ of the system. Preferably, we would like to work in the thermodynamic limit $V\to\infty$. For compact systems, one can easily construct a set of variational ansatz states by taking the span of the ground state and a few of the excited states of a nearby free Hamiltonian that can be diagonalized exactly. For extended systems, this approach will fail because the excited states will not be extensive and can thus not contribute to the energy expectation value. We thus end up trying to apply the variational method by using the ground state of a free theory --- a \emph{Gaussian} state --- as variational ansatz, which is equivalent to mean field theory. If we do try to devise wave functionals which are extensive and non-Gaussian, we meet Feynman's third objection: we still have to compute functional integrals in order to calculate the energy expectation value. This is as difficult as calculating the path integral describing the QFT, but in one dimension less, since time does not appear in the Hamiltonian framework. For non-Gaussian states, Feynman believed that it is impossible to accurately calculate expectation values, as he considered perturbation theory the only means possible to compute these integrals. The resulting errors have a strong influence on the optimal state $\ket{\Psi(z^\ast)}$ obtained by applying the variational method and thus on observable quantities derived from it. 

While some progress into non-Gaussian variational methods for relativistic QFTs and gauge theories have been developed since Feynman's original lecture \cite{othernongaussian}, we hope that we can contribute with a new approach. 

\section{A new variational ansatz: continuous matrix product states}
Feynman's second and third arguments are not restricted to relativistic theories, and should thus be equally valid for all extended QFTs and even quantum lattice systems. His statement follows from his sole consideration of perturbation theory as the only possible method to calculate expectation values of non-Gaussian states and expresses a ``lack of imagination''. Since 1988 \cite{AKLT} we have witnessed a stream of results and variational ansatzes for low-dimensional quantum lattice systems. The most striking development was the introduction of the DMRG \cite{White}, which is --- in retrospect --- a variational method within the class of \emph{matrix product states} (MPS) \cite{MPS, reviewMPS}. The basis of the success of the DMRG is the fact that MPS correctly capture the amount of \emph{quantum entanglement} in gapped one-dimensional lattice systems. A better understanding of the behavior of entanglement in ground states of short-ranged Hamiltonians --- namely the fact that the entanglement entropy of a region scales as the boundary of that region, with at most logarithmic violations for critical systems --- has led to the conclusion that ground states of such Hamiltonians live in a very small corner of the Hilbert space. This insight was crucial in the development of new variational wave functions for strongly-interacting quantum systems that live precisely in this corner and have the required scaling of entanglement entropy. The natural generalization of MPS to systems satisfying an area law in higher dimensions are the projected entangled pair states (PEPS), whereas the logarithmic violations of the boundary law in $(1+1)$ dimensional critical systems can be captured by the multi-scale entanglement renormalization ansatz (MERA) \cite{MPStoMERA}. These wave functions go well beyond Gaussian trial states and allow the accurate and efficient calculation of observable quantities. Thus Feynman's objections can already be regarded as having been addressed, in the case where MPS are applied to relativistic QFTs in conjunction with a lattice regulator \cite{mpsforqft}.

These developments have culminated with the introduction of a new variational wave functional for $(1+1)$-dimensional QFTs: the \emph{continuous matrix product state} (cMPS) \cite{cMPS,HQS}. They are obtained as the continuum limit of a certain subclass of MPS and provide a variational class of non-Gaussian wave functionals directly for quantum fields, removing the need for a lattice regulator. Compelling evidence that cMPS provide a powerful description of the quantum fluctuations of quantum fields has been presented in \cite{cMPS, maruyama:2010a}. So far, cMPS have been restricted to the approximation of ground states of non-relativistic theories. In this presentation we argue that we can apply the cMPS wave functional to relativistic QFTs and illustrate how to overcome Feynman's first argument, which is inextricably connected with any attempt to apply the variational approach to relativistic QFTs. As for all variational approaches, we work in the Hamiltonian framework and explicitly specify the ground state wave functional. While this is not a common approach to tackle relativistic QFTs, the Schr\"odinger formalism for relativistic QFTs is well-established (see \cite{jackiw} and references therein).

A final remark is in order before introducing the cMPS ansatz. In his closing remarks, Feynman speculated how best to overcome his second reservation and predicted that it should be possible to describe a global field state using a reduced set of local parameters. Feynman foresaw the role of the \emph{density matrix} in such a description. It turns out that the \emph{density matrix} in DMRG has precisely the properties envisaged by Feynman: it yields a local parameterization of the global properties of a state which is living on the boundary of the region of interest. CMPS inherit this as they possess a key \emph{holographic property}: they are parameterized by the (non-equilibrium) dynamics of an auxiliary system --- which we could call a \emph{boundary field theory} --- of one lower geometric dimension \cite{cMPS, HQS}. For $(1+1)$ dimensional QFTs, the auxiliary system is zero-dimensional (the Hilbert space of the auxiliary system can even be chosen to be finite dimensional) and is thus exactly solvable. While this is no longer true in higher-dimensional generalizations, we foresee that approximations similar to those used in the contraction of PEPSs will still allow for an accurate calculation of expectation values.

While Feynman's arguments are valid both for bosonic and fermionic theories, we focus on fermionic theories as these are naturally formulated in terms of creation and annihilation operators, which is compatible with the cMPS formalism. We can define the fermionic cMPS class as:
\begin{equation*}
\ket{\Psi}=\mathrm{Tr}_\text{aux} \left[\mathcal{P}\mathrm{e}^{\int_{-\infty}^{+\infty}\mathrm{d}x\,Q\otimes  \openone+ \sum_\alpha R_\alpha\otimes \hat{\psi}_ \alpha ^\dagger(x)}\right]\ket{\Omega},\label{eq:ansatz}\vspace{-2ex}
\end{equation*}
where $\hat{\psi}_\alpha^\dagger(x)$ are field operators creating fermions of type $\alpha$ at position $x$ with anticommutation relations $\{\psi^\dagger_\alpha(x),\psi^\dagger_\beta(y)\}=0$ and $\{\psi^\dagger_\alpha(x),\psi_\beta(y)\}=\delta_{\alpha,\beta}\delta(x-y)$, $Q$ and $R_\alpha$ are $D\times D$ matrices acting on the auxiliary system, $\mathrm{Tr}_\text{aux}$ denotes a partial trace over the auxiliary system, and $\mathcal{P}\mathrm{e}$ denotes the path ordered exponential. The matrices $Q$ and $R_{\alpha}$ contain the variational parameters and can be position dependent, but we focus on a translational-invariant setting where they are not. A derivation of the required algorithmic rules for calculating expectation values of cMPS can be found in \cite{cMPS,HQS} and we only highlight differences resulting from the anticommutation relations of the fermionic field operators. In the relativistic scenario, the two field operators $\hat{\psi}_\alpha^\dagger$ ($\alpha=1,2$) represent the two components of the Dirac spinor. The state $\ket{\Psi}$ approximates the ground state of a relativistic QFT by acting with the field creation operators on the state $\ket{\Omega}$, for which all levels are empty ($\hat{\psi}_\alpha \ket{\Omega}=0$). For free Dirac fermions, the path-ordered exponential should thus fill the Dirac sea.

\section{A natural cutoff}
Let's now describe the physical properties of the fermionic cMPS variational class. It is a non-Gaussian class that is both extensive --- note the action of the creation operator inside the exponential --- and allows the exact evaluation of the expectation values of local operators; \textit{e.g.}\ we obtain (we henceforth use the summation convention on repeated indices):
\begin{equation*}
-\frac{\mathrm{i}}{2}\braket{\chi|\hat{\psi}^\dagger \alpha^{x} \frac{\mathrm{d}\hat{\psi}}{\mathrm{d}x}(x)|\chi}+\frac{\mathrm{i}}{2}\braket{\chi| \frac{\mathrm{d}\hat{\psi}^{\dagger}}{\mathrm{d}x}(x)\alpha^{x}\hat{\psi}(x)|\chi}  = \textrm{Im} [\sigma^y_{\alpha\beta}\bra{l} [Q, R_\alpha]_-\otimes \overline{R}_\beta\ket{r}],
\end{equation*}
for the kinetic energy density, where, in order to obtain real coefficients, we have chosen the convention $\alpha^x=\sigma^y$ and $\beta=\sigma^z$ for the Dirac matrices. The $D^2$ component vectors $\bra{l}$ and $\ket{r}$ are, respectively, the left- and right-eigenvectors of the \emph{transfer matrix} $T=Q\otimes \openone + \openone \otimes \overline{Q} + R_\alpha \otimes \overline{R}_\alpha$, corresponding to eigenvalue zero \cite{cMPS,HQS}. We focus on the kinetic energy density as it is the dominant term in the UV region, which is the region responsible for divergences and for Feynman's first criticism. As long as the $D\times D$ matrices $Q$ and $R_\alpha$ have finite entries this expression will be finite and is thus regularized.

A better understanding of this regularization is gained by looking at the momentum occupation in a cMPS: $\braket{\chi|\hat{\psi}^\dagger_\alpha(k)\hat{\psi}_\beta(k^\prime)|\chi}=\delta(k-k^\prime) n_{\alpha,\beta}(k)$ \cite{TIcomment}. The momentum occupation number $n_{\alpha,\beta}(k)$ is the Fourier transform of $C_{\alpha\beta}(x)$, where
\begin{equation*}
C_{\alpha,\beta}(x)=\theta(-x)\braket{l| (\openone \otimes \overline{R}_\alpha) \mathrm{e}^{x\widetilde{T} }(R_\beta\otimes \openone)|r}+\theta(x)\braket{l|  (R_\beta\otimes \openone)\mathrm{e}^{x\widetilde{T}}(\openone \otimes \overline{R}_\alpha)|r}\label{eq:correlator}
\end{equation*}
and $\theta(x)$ the Heaviside function and $\widetilde{T}=Q\otimes \openone + \openone \otimes \overline{Q} - R_\alpha \otimes \overline{R}_\alpha$, where the last minus sign originates from the Fermi statistics of the particles. There will not be any disconnected contribution, as we require $\braket{\chi |\hat{\psi}_\alpha|\chi}=0$. The behavior of $n_{\alpha,\beta}(k)$ for large $k$ is determined by the continuity and differentiability of $C_{\alpha,\beta}$, in particular around $x=0$, which is the only point where differentiability of the expression above is not trivially guaranteed. Since $C_{\alpha,\beta}(x)$ is a continuous function, its Fourier transform decays as $n_{\alpha\beta}(k) \leq \mathcal{O}(k^{-2})$ for $|k| \to \infty$. Continuity of the derivative of $C_{\alpha,\beta}(x)$ at $x=0$ requires $\braket{l|\left\{R_\beta,R_\gamma\right\}\otimes \left\{\overline{R}_\alpha,\overline{R}_\gamma\right\} |r}=0$ ($\forall \alpha, \beta$), which is satisfied by choosing all matrices $R_\alpha$ nilpotent and anticommuting. The second derivative of $C_{\alpha,\beta}(x)$ at $x=0$ is then automatically continuous, from which one can conclude that $n_{\alpha,\beta}(k) \leq \mathcal{O}(k^{-4})$ for $|k| \to \infty$. While a faster decrease of the momentum occupation number imposes additional constraints on the matrices $Q$ and $R_\alpha$ the current behavior already ensures a finite kinetic energy. The region in momentum space where the $k^{-4}$ decay behavior sets in defines a soft momentum cutoff $\Lambda$. 

\section{Curing the sensitivity to high frequencies}
We can now investigate how Feynman's first objection manifests itself for the cMPS ansatz. The problem is situated in a cMPS's ability to describe a scale transformation $x\mapsto cx$ ($c>0$) by an equivalent transformation $Q' = cQ$ and $R_\alpha' = \sqrt{c}R_\alpha$. Since this transformation does not change $\bra{l}$ and $\ket{r}$, the kinetic energy per unit length will simply be multiplied by a factor $c^2$. In renormalizable theories, the kinetic energy has the highest scaling dimension, together with other terms with dimensionless coupling constants. These are thus the dominant terms in the UV region. However, in contrast to the non-relativistic case, the relativistic kinetic energy is not a positive definite operator and can acquire a negative expectation value. If $|\chi\rangle$ is a cMPS for which the sum of terms with highest scaling dimension has a negative energy expectation value, then the total ground state energy can always be decreased by a scale transformation with $c$ sufficiently large. Our variational method will thus try to push $c\to \infty$, in order to approximate the divergent (kinetic) energy of the exact solution. Under such a transformation, the momentum occupation changes to $n^\prime_{\alpha,\beta}(k)=n_{\alpha,\beta}(k/c)$ and the intrinsic cutoff determined by $n^\prime$ is given by $\Lambda^\prime=c\Lambda$. 

\begin{figure}
\centering
\includegraphics[width=0.7\textwidth]{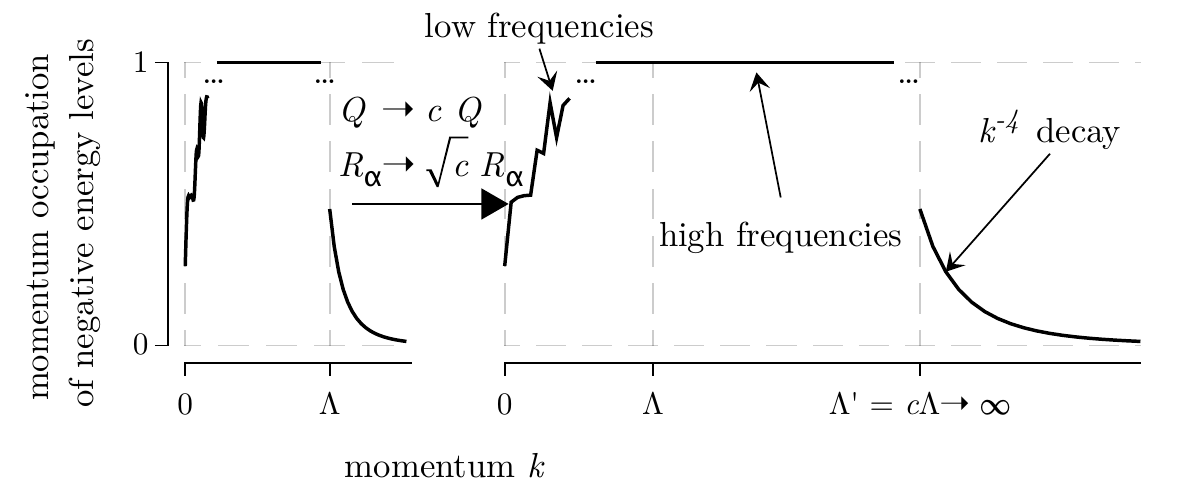}
\caption{Hypothetical momentum distribution of an optimal cMPS for a free fermionic theory: high-frequency degrees of freedom are well-approximated up to a cutoff $\Lambda$, after which the momentum occupation decays as $k^{-4}$. Also shown is the effect of a scale transformation.\label{fig:sketch}}
\end{figure}

This change of scale will be accompanied by a worse description of the low frequency region, as predicted by Feynman. The precise underlying cause for this effect in our variational class is that a cMPS can only accurately describe states with a finite amount of entanglement. The maximal entanglement entropy in a one-dimensional system with energy gap $\Delta$ and energy cutoff $\Lambda$ will  roughly be given by $S\sim \log(\Lambda/\Delta)$, and a cMPS with $D$ proportional to $\mathcal{O}(\exp(S))$ should suffice to provide a good description \cite{arealaws}. If $D$ is too low to obtain a good approximation of the exact ground state, the variational method will make compromises in that part of the frequency spectrum that contributes least to the ground state energy, \textit{i.e.}\ the low-frequency region. In non-relativistic systems, the cutoff is set by the particle density or thus by the chemical potential. But in a relativistic Hamiltonian, there is no physical cutoff and we only have the intrinsic momentum cutoff $\Lambda$ of the cMPS. If we start from a cMPS with negative energy expectation value, the variational method can quickly lower the energy by shifting the cutoff to $\Lambda'=c\Lambda$ with $c \to \infty$. As $c$ goes to infinity, all low-energy modes will eventually fall into the region that is poorly described and the description of any observable quantity will be completely wrong for every finite value of the bond dimension $D$. This is schematically illustrated in Fig.~\ref{fig:sketch}.

A solution is now straightforward as we can prevent $c$ from running to infinity by imposing a constraint on the matrices $Q$ and $R_\alpha$: since $Q$ has the dimension of momentum and $R_\alpha$ has the dimension of the square root of a momentum, constraining the \emph{norm} of $Q$ and $R_\alpha$ prevents $c$ from running and regularizes the resulting theory by introducing a scale, i.e.\ a dimensionful parameter, into the system, similar to what happens in analytical regularization techniques or lattice regularization. In the sequel, we will constrain the norm of the commutator $[Q,R_\alpha]$ by fixing the expectation value of $(\mathrm{d}\hat{\psi}^\dagger/\mathrm{d} x) (\mathrm{d} \hat{\psi}/\mathrm{d} x)$ \cite{cutoffcomment}. Hereto, we add this term to the Hamiltonian with a Lagrange multiplier $1/\Lambda$, \textit{i.e.} $\hat{H}_\text{cutoff}=\Lambda^{-1} \int\mathrm{d} x\, (\mathrm{d}\hat{\psi}^\dagger/\mathrm{d} x) (\mathrm{d} \hat{\psi}/\mathrm{d} x)$. This apparently arbitrary choice is motivated by the requirement that the constraint needs to penalize high values of the momentum $k$, to which $[Q,R_\alpha]$ is related by the calculational rules of cMPS. $H_\text{cutoff}$ will give a $k^2$ contribution in momentum space, which is low enough to ensure a finite result in combination with a momentum occupation that decays as $k^{-4}$. It is, however, strong enough to penalize high frequency modes, even the ones that give a contribution $-|k|$ to the (kinetic) energy. Put differently, it is a positive definite term with a higher scaling dimension than the relativistic kinetic energy. As such, it is non-renormalizable, which by means of the renormalization group indicates that it will be irrelevant for the description of the low-frequency modes and cannot strongly influence the expectation value of observable quantities. Note that it does respect the chiral symmetry of the kinetic energy term. It does of course break relativistic invariance, which is inevitable when introducing a momentum cutoff in a Hamiltonian framework. We expect that any other norm constraint with similar properties and respecting the symmetries of the system should also work.

\section{Application 1: free Dirac fermions}

\begin{wrapfigure}{r}{0.5\textwidth}
\centering
\includegraphics[width=0.5\textwidth]{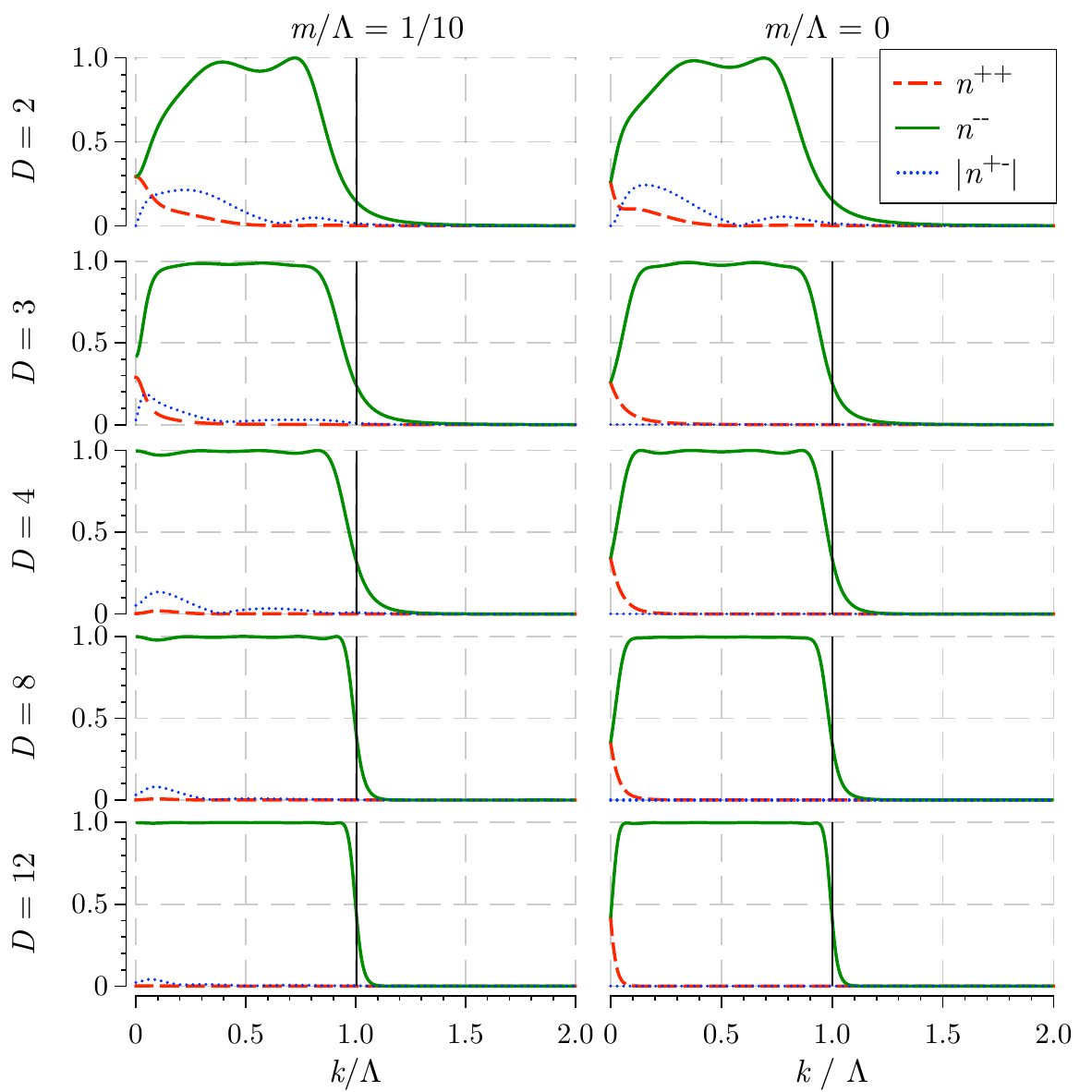}
\caption{Momentum occupation of the antiparticle levels $n^{--}(k)$, the particle levels $n^{++}(k)$ and the mixing $|n^{+-}(k)|$ in a cMPS approximation of the Dirac field with mass $m$. The auxiliary space of the cMPS is $\mathbb{C}^2\otimes \mathbb{C}^2\otimes \mathbb{C}^D$. The vertical line indicates the position of the exact cutoff.\label{fig:dirac}}
\end{wrapfigure}
We now illustrate our arguments by applying them to relativistic fermion models. As a benchmark, we first consider free Dirac fermions with Hamiltonian density
\begin{equation*}
\begin{split}
\hat{h}_{D} =& -\frac{\mathrm{i}}{2} \hat{\psi}^\dagger(x) \sigma^y \frac{\mathrm{d}\hat{\psi}}{\mathrm{d} x}(x) + \frac{\mathrm{i}}{2} \frac{\mathrm{d}\hat{\psi}^\dagger}{\mathrm{d} x}(x) \sigma^y \hat{\psi}(x)\\
&+ m \hat{\psi}^\dagger(x) \sigma^z \hat{\psi}(x),
\end{split}
\end{equation*}
with Dirac matrices chosen as described above, and $m$ the fermion mass. In the exact ground state of $\hat{H}_{D}+\hat{H}_\text{cutoff}$, this term will actually introduce a sharp cutoff at $k_\text{cutoff}=\Lambda (1/2 + (1/4+m^2/\Lambda^2)^{1/2})^{1/2}$, which is equal to $\Lambda$ up to corrections of $\mathcal{O}(m^2/\Lambda^2)$. The cMPS ansatz will not be able to reproduce this sharp cutoff because it decays as $k^{-4}$. Indeed, this cutoff is not expected to be reproduced very well, because the new hamiltonian is gapless at $k=\pm k_\text{cutoff}$. However, this is not a problem, as we do not expect these high-frequency modes to influence physical properties. Note that both zeros in the dispersion relation occur at physically different momenta and do thus not result in fermion doubling.

Since we do not aim at reproducing the exact solution in the high-frequency regime, we can not compare the corresponding energy as a measure of the accuracy of our solution. Instead, we have calculated the momentum occupation of the exact positive (particle) and negative (antiparticle) levels according to the definitions
$\braket{\operator{a}^\dagger(k) \operator{a}(k^\prime)}=\delta(k^\prime-k) n^{++}(k)$, $\braket{\operator{b}^\dagger(k) \operator{b}(k^\prime)}=\delta(k^\prime-k) n^{--}(k)$, $\braket{\operator{a}^\dagger(k) \operator{b}(k^\prime)}=\delta(k^\prime-k) n^{+-}(k)$, with $\operator{a}$ ($\operator{b}$) the annihilator (creator) of particles (antiparticles). The results corresponding to the optimal cMPS are shown in Fig.~\ref{fig:dirac}. The exact solution has the Dirac sea filled ($n^{--}(k)=1$) all the way up to $k_\text{cutoff}$, after which $n^{--}(k)=0$ for $|k|>k_\text{cutoff}$, and $n^{++}(k)=n^{+-}(k)=0$, $\forall k$. These results were obtained using the cMPS ansatz where $Q$ and $R_\alpha$ act on an auxiliary Hilbert space $\mathbb{C}^{2}\otimes \mathbb{C}^{2}\otimes \mathbb{C}^D$, where the first two two-dimensional Hilbert spaces accommodate auxiliary fermions which are used to impose the anticommutation relations on $R_\alpha$. It is clear from these results that the low-energy behavior is approximated very well for the massive Dirac theory, and the accuracy greatly increases by increasing $D$. As anticipated, the cutoff behavior is approximated less well. In the case $m=0$ the theory is critical and the low-energy behavior is also approximated less well. This result is familiar from MPS solutions for gapless lattice models. However, from the fact that $|n^{+-}|\approx 0$ for $m=0$, we see that the algorithm automatically converges to a cMPS respecting chiral symmetry, except at $D=2$.

\begin{wrapfigure}{r}{0.5\textwidth}
\begin{minipage}{0.5\textwidth}
\centering
\includegraphics[width=\textwidth]{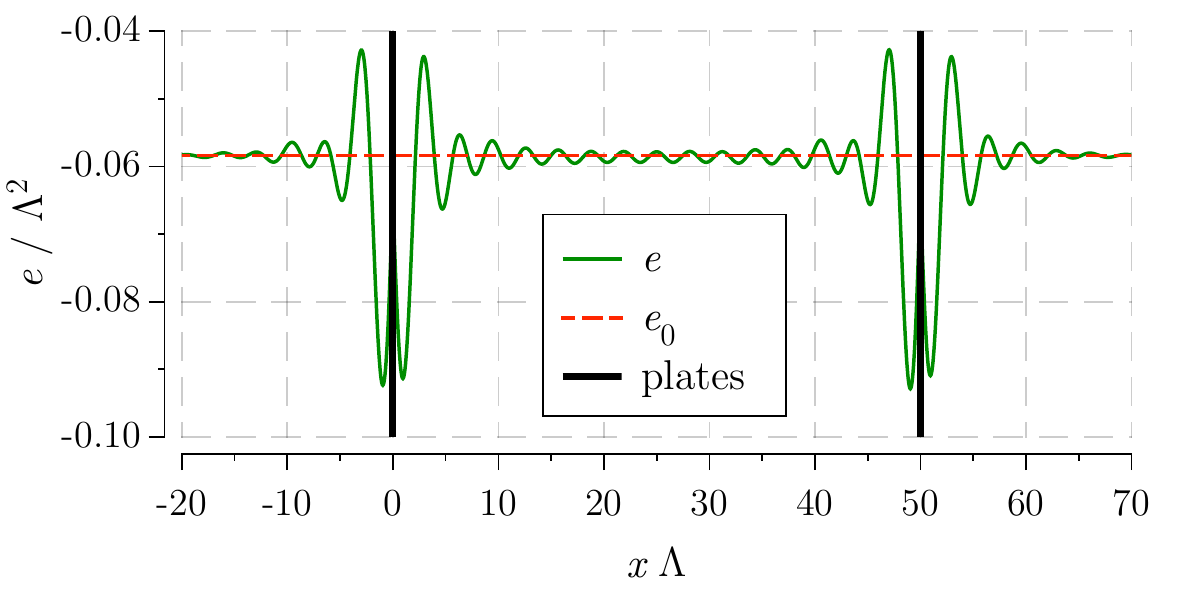}
\caption{Energy density $e$ for Dirac fermions with $m/\Lambda=1/10$ in a system where `plates' are present at position $x=0$ and $x=50/\Lambda$. These plates enforce the bag-model boundary conditions. The ground state energy density $e_0$ in the infinite vacuum is plotted for comparison.\label{fig:casimirsingle}}
\end{minipage}
\vspace{1ex}

\begin{minipage}{0.5\textwidth}
\centering
\includegraphics[width=\textwidth]{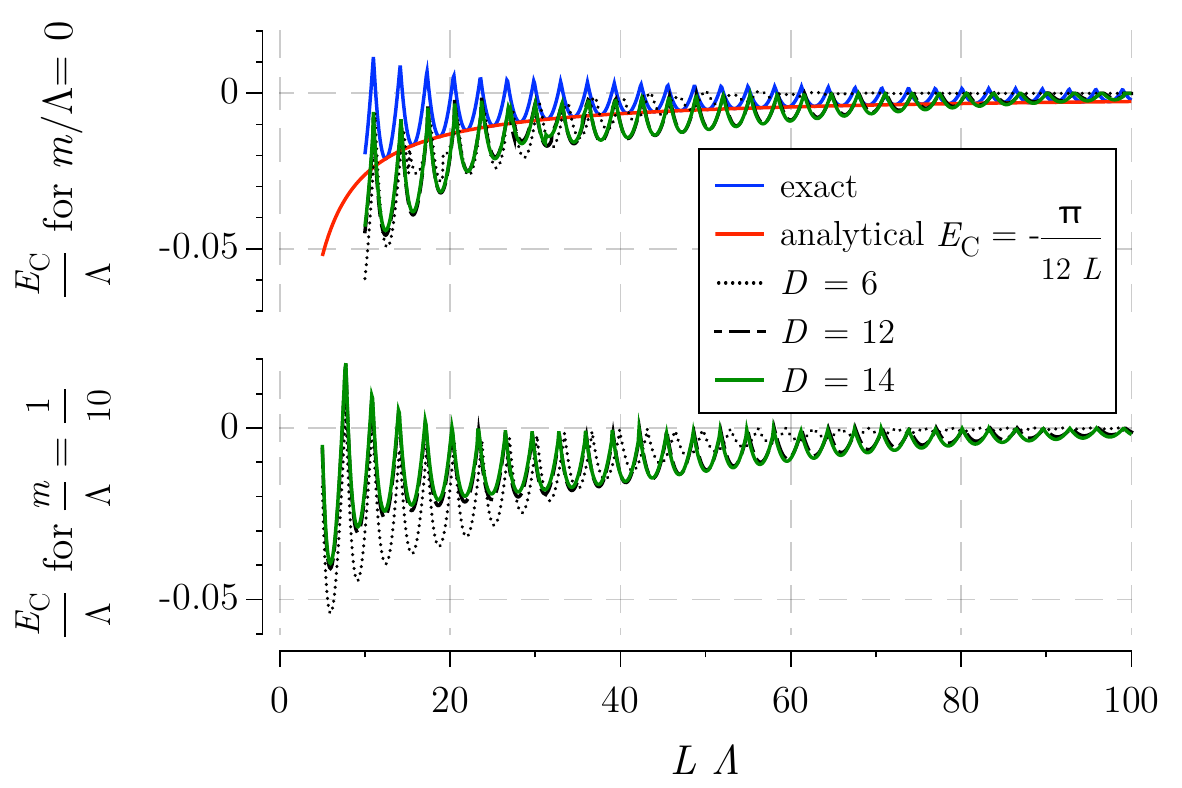}
\caption{The total Casimir energy as a function of the distance $L$ between the `plates'. For $m=0$, the exact Casimir energy, both in our model with cutoff, and analytically through zeta function regularization, is also displayed. \label{fig:casimirtotal}}
\end{minipage}
\end{wrapfigure}

To give a non-trivial example of what can be done with our approach we have also calculated the Casimir energy of the massive Dirac field. We simply recycle the matrices $Q$ and $R_\alpha$ from the simulation above, and add suitable operators $B(x)$ to the ansatz at the location of the `plates' or defects ($x=0$ and $x=L$), which impose the correct boundary conditions. The boundary conditions will only fix a part of these operators, the remaining elements can be used to optimize the energy with fixed $Q$ and $R_\alpha$. All boundary effects can be incorporated in the boundary operators $B$. We used the typical bag-model boundary conditions \cite{milton}. In Fig.~\ref{fig:casimirsingle} we show the energy density for a particular configuration of `plates' in the $(1+1)$ dimensional free-fermion model. A clear manifestation of the Fermi surface at a finite momentum $k_\text{cutoff}$ is present in the form of Friedel oscillations. The Casimir energy $E_C$ as a function of the distance $L$ between the plates is plotted in Fig.~\ref{fig:casimirtotal}. The presence of the momentum cutoff, and thus of the finite particle density, also introduces a strong oscillatory behavior in $E_C(L)$, which was already observed in studies of the interaction energy between defects in one-dimensional quantum liquids \cite{casimirquantumliquids}. Local minima correspond to values of $L$ where the number of allowed modes is such that the density of fermions between the plates is exactly equal to the density of fermions outside the plates. The sharp maxima appear when this condition is most strongly violated. In the limit $k_\text{cutoff}\to \infty$, the density of fermions is infinite, both in between and outside the plates, and the equal density condition is always satisfied. So the physical Casimir energy, which is expected to be cutoff independent, can be found by the envelope of the local minima. This is illustrated for the $m=0$ case, where the exact solution in our model is compared to the value of the Casimir energy for massless Dirac fermions in $(1+1)$ dimensions, as calculated with zeta-function regularization, i.e.\ $E_\text{C}(L)=-\frac{\pi}{12 L}$. Note that the Casimir energy will always have an asymptotic exponential decay in cMPS calculations, but that it can be well approximated at intermediate values of $L$.

Because the Casimir energy is a \emph{difference} of energies approximate results can be lower than the exact solutions. This is clearly the case in  Fig.~\ref{fig:casimirtotal}. We attribute this effect to the fact that the additional degrees of freedom present in the boundary vectors allow one to further optimize the energy in their immediate vicinity. The qualitative behavior of the energy is already reproduced by this simple approach.

\section{Application 2: the Gross-Neveu model}
\begin{figure}
\centering
\includegraphics[width=0.6\textwidth]{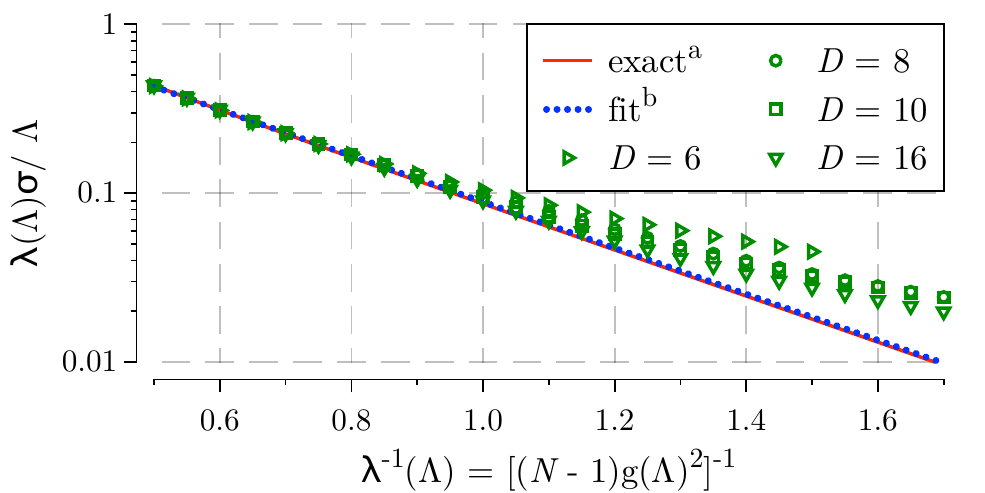}
\caption{Expectation value of $\sigma=\bra{\chi}\hat{\psi}^\dagger\sigma^z \hat{\psi}\ket{\chi}$ in the Gross-Neveu model as function of $\lambda(\Lambda)$ for $N=\infty$. A fit of the form $c_1e^{-c_2/\lambda}$ to the numerical results for $\lambda^{-1} \leq 1$ at $D=16$ results in $c_2 = 3.142^{+0.047}_{-0.047}$ and $c_1=2.057^{+0.074}_{-0.072}$, to be compared to the exact values $c_{1}=2$ and $c_{2}=\pi$ (see main text).}
\label{fig:grossneveu}
\end{figure}

As a final proof of principle, we study a theory with interactions. One of the most important models for one-dimensional relativistic fermions is the Gross-Neveu model, as it shares many features with QCD \cite{grossneveu}, including, asymptotic freedom and spontaneous breaking of chiral symmetry. The hamiltonian density for the $N$-flavor Gross-Neveu model is given by
\begin{align*}
\hat{h}_{GN} = -\frac{\mathrm{i}}{2} \hat{\psi}^\dagger_a \sigma^y \frac{\mathrm{d}\hat{\psi}_a}{\mathrm{d} x} + \frac{\mathrm{i}}{2} \frac{\mathrm{d}\hat{\psi}^\dagger_{a}}{\mathrm{d} x}(x) \sigma^y \hat{\psi}_{a}(x)- \frac{g^2}{2}: (\hat{\psi}_a^\dag \sigma^z \hat{\psi}_a)^2 :,
\end{align*}
where the $x$-dependence of the field operators has been omitted for brevity and there is an implied summation over the flavor index $a=1, 2, \ldots, N$. One must not forget to apply normal ordering when deriving an interacting Hamiltonian from the relativistic path integral, which is a coherent-state path integral for fermionic theories. As a variational ansatz we employ a product state of cMPS states across the different fermion flavors. Because the exact ground state has $\mathcal{S}_N$ flavor symmetry (and actually $\mathrm{O}(2N)$ symmetry), the nearest product state should also be invariant under $\mathcal{S}_N$ \cite{symproductstate}. We can thus use the same cMPS for every flavor. This amounts to a Hartree-Fock approximation of the theory, where the self-interaction of the flavor is treated exactly, and the self-consistent mean-field approach is only applied to the interactions between different flavors. We add the same cutoff term $\operator{H}_{\text{cutoff}}$ to the Hamiltonian for every fermion flavor, so as to respect the flavor symmetry. Since this term introduces our regularization parameter $\Lambda$, we know that the coupling constant $g$ will have to depend on $\Lambda$ in order to have a consistent theory. In the $N\to \infty$ limit, we can solve this problem exactly, and we obtain the well-known result for $\sigma = \braket{\chi|\hat{\psi}^\dagger \sigma^z \hat{\psi} | \chi}$
\begin{equation*}
\frac{\pi}{\lambda}=\int_0^{k_\text{cutoff}}\frac{\mathrm{d} k}{\sqrt{\lambda^2 \sigma^2+k^2}}\quad \Rightarrow \quad |\lambda\sigma| \approx 2 \Lambda \mathrm{e}^{-\frac{\pi}{\lambda(\Lambda)}}\label{grossneveuresult}
\end{equation*}
where $k_\text{cutoff}\approx \Lambda$ if $|\lambda\sigma| \ll \Lambda$. This indicates that the cutoff fixing term $\hat{H}_\text{cutoff}$ has no effect other then what it is meant to do, i.e.\ introducing a cutoff. With the current Hartree-Fock ansatz based on cMPS, we can calculate an approximation for any $\lambda$ and $N$. In principle, we can describe the complete wave function of all $N$ flavors with a single cMPS, but this is computationally more demanding as the dimension of the auxiliary space needs to grow exponentially with the number of flavors. Numerical results with the mean-field approach are illustrated in Fig.~\ref{fig:grossneveu}. At strong coupling ($\lambda>1$) they agree very well with the exact result. The discrepancies between the exact solution and the cMPS approximation for $N=\infty$ are clearly finite-$D$ effects. They become more pronounced as $\lambda \sigma/\Lambda$ gets smaller, since $\lambda\sigma$ is exactly equal to the mass gap in the $N=\infty$ limit.

\section{Conclusion}

We have developed an extension of the cMPS variational class appropriate for fermionic \mbox{$(1+1)$}-dimensional relativistic field theories. Since cMPS have a built-in cutoff, they offer a new way to regularize quantum field theories. We have explained how to ensure that cMPS do not suffer from Feynman's objections concerning the application of the variational principle. Additionally, our approach is free from fermion doubling and sign problems. We have demonstrated the applicability of our variational approach by reproducing the known results for free Dirac fermions and provided two nontrivial applications of our method to the Gross-Neveu model, where we observe chiral symmetry breaking and absolute scaling of the chiral parameter, and the Casimir effect, where we are able to reproduce the qualitative features of the Casimir energy.

\section*{Acknowledgements}
J.H.\ and H.V.\ would like to thank the hospitality of R.~Bertlmann and F.V.\ at the University of Vienna. T.J.O.\ is grateful to J.~Eisert for helpful conversations. Work suported by Research Foundation Flanders (JH), SFB projects, FoQuS and ViCoM, EU projects Quevadis, ERC grant QUERG and DFG-FG635.


\begin{thebibliography}{99}
\bibitem{TIcomment}
	The delta function $\delta(k-k')$ arises because $|\Psi\rangle$ is translation invariant.
\bibitem{cutoffcomment}
In terms of the matrices $Q$ and $R_\alpha$, the expectation value of $\hat{h}_\text{cutoff}$ is given by $\braket{l| [Q,R_\alpha]\otimes [\overline{Q},\overline{R}_\alpha]|r} = \mathrm{Tr}\left([Q,R_\alpha]\rho_0 [Q,R_\alpha]^\dagger\right)$ with $\rho_0$ the positive definite (density) matrix associated with $\ket{r}$. This expression is thus a valid norm for $[Q,R_\alpha]$.
\end{thebibliography}
\end{document}